\documentclass[aps,prl,twocolumn,superscriptaddress,groupedaddress]{revtex4}  
\usepackage{graphicx,comment,float,braket,textcomp,amsmath,lipsum,color,soul}  
\usepackage{dcolumn}   
\usepackage{bm}        
\usepackage{amssymb}   

\begin{document}

\title{Excitation-Dependent Features and Artifacts in 2-D Terahertz Spectroscopy}

\author{Albert Liu}
\email{aliu1@bnl.gov} 
\affiliation{Condensed Matter Physics and Materials Science Division, Brookhaven National Laboratory, Upton, New York, 11973 USA}

\author{Ankit Disa} 
\affiliation{School of Applied and Engineering Physics, Cornell University, Ithaca, New York, 14850 USA}

\vskip 0.3cm

\date{\today}




\begin{abstract}
    Recently, two-dimensional terahertz spectroscopy (2DTS) has attracted increasing attention for studying complex solids. A number of recent studies have applied 2DTS either with long pulses or away from any material resonances, situations that yield unconventional 2DTS spectra that are often difficult to interpret. Here, we clarify the generic origins of observed spectral features by examining 2DTS spectra of ZnTe, a model system with a featureless optical susceptibility at low terahertz frequencies. These results also reveal possible artifacts that may arise from electro-optic sampling in collinear 2DTS experiments, including the observation of spurious rectified or second harmonic signals.
\end{abstract}

\maketitle

\section{Introduction}
Two-dimensional optical spectroscopy \cite{Cundiff2013,MDCS_Book} and its predecessor two-dimensional nuclear magnetic resonance (NMR) \cite{ernst1990principles} are powerful techniques that have found widespread application in both industrial application and fundamental science. Recently, advances in laser technology have extended these multi-dimensional spectroscopies into terahertz frequencies in which many fundamental excitations of condensed matter are found. So-called 2-D THz spectroscopy (2DTS) \cite{Johnson2019,Lu2019,Reimann2021} has been demonstrated in a variety of material systems, such as molecular gases \cite{Lu2016,Gao2022}, semiconductors \cite{Maag2016,Somma2016,Houver2019,Mahmood2021}, ferroics \cite{Pal2021,Lin2022,Zhang2024}, superconductors \cite{Luo2023,Liu_2023_echo,Kim2024}, and even topological materials \cite{Blank2023}.

Two-dimensional optical spectroscopy and two-dimensional NMR are most often performed and interpreted in the impulsive limit \cite{Mukamel1999_Book,HammZanni2012}, in which the measured nonlinear wave-mixing signal directly reflects a microscopic optical response function with only a scalar dependence on the excitation field amplitudes $E_A$ and $E_B$ \cite{Salvador2024}. As shown in Figure~\ref{Fig1}(a), in the time-domain this limit corresponds to excitation pulse durations far shorter than any relevant dephasing times, while in the frequency-domain this limit corresponds to an excitation bandwidth far broader than the spectral content of the relevant optical response function. 

These conditions, however, are increasingly violated in 2DTS experiments, due to both limitations of terahertz light sources and the relevance of low-energy continua in the physics of solid-state materials. For example, as shown in Figure~\ref{Fig1}(b), excitation pulse durations may be comparable to system dephasing times which corresponds to excitation spectral bandwidths comparable or greater than resonance linewidths. In addition, as shown in Figure~\ref{Fig1}(c), the excitation bandwidth may not be centered on any resonance, probing only a featureless continuum. Though distinct in origin, these two cases both result in a nonlinear signal that is generated only during pulse overlap and a two-dimensional spectrum that is shaped primarily by the excitation pulse spectrum.

Here, we aim to clarify the generic spectral features of non-resonant (and non-impulsive) 2DTS experiments. We present 2DTS spectra of ZnTe, a model system with a featureless optical susceptibility in the few-THz frequency regime, which exhibit characteristic features due to both second- and third-order optical nonlinearities that are well-reproduced by simulation. As will be elucidated below, these results also reveal artifacts that may arise in 2DTS experiments due to a collinear detection setup.

\section{Experimental details}
Figure~\ref{Fig2}(a) illustrates our experimental setup, which emulates a typical 2TDS experiment in a reflection geometry. Two terahertz excitation pulses $E_A$ and $E_B$ are generated by optical rectification in two separate organic crystals DAST (at a 1 kHz repetition rate), which are combined collinearly using an initial silicon beamsplitter. These pulses then propagate through a second silicon beamsplitter to the sample position, and their subsequent reflections then propagate to the electro-optic sampling (EOS) position where they are combined with an 800 nm probe pulse on a ZnTe crystal of 150 $\mu$m thickness. There, we estimate peak fields on the order of tens of kV/cm per pulse. In the usual case, the ZnTe would act as an EOS detector for the nonlinear signal generated by the mixing of $E_A$ and $E_B$ at the sample position, but here we examine the characteristic 2TDS responses of the ZnTe itself to the non-resonant excitation from the incident fields.

\begin{figure}[t]
    \centering
    \includegraphics[width=0.45\textwidth]{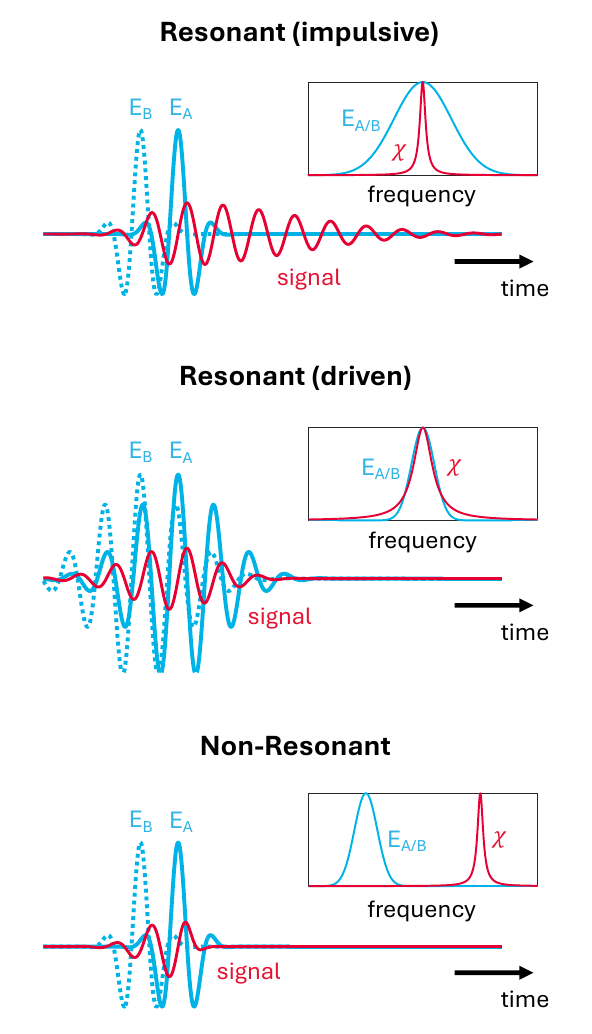}
    \caption{Three regimes of nonlinear signal generation by two excitation pulses $E_A$ and $E_B$ in 2DTS. (a) Impulsive excitation of a resonance $\chi$, in which excitation pulse durations are far shorter than relevant dephasing times. (b) Driven excitation of the resonance, in which pulse durations are comparable to relevant dephasing times. (c) Off-resonant excitation, in which the excitation spectrum does not overlap with any material resonance. Insets depict corresponding spectral overlap of the excitation spectra $E_{A/B}$ and optical response $\chi$.}
    \label{Fig1}
\end{figure}

With its first optical phonon resonance found at 5.3 THz \cite{Wu1996}, ZnTe offers a relatively featureless optical susceptibility toward lower frequencies \cite{Hattori1973}. To examine the non-resonant excitation condition shown in Figure~\ref{Fig1}(c), we therefore place a gold mirror at the sample position so that the primary nonlinearity driven by $E_A$ and $E_B$ is that of the ZnTe electro-optic sampling crystal itself. A standard double-chopping scheme is used in which pulses A and B are modulated at 1/2 kHz and 1/3 kHz respectively, resulting in a nonlinear signal that can be lock-in detected at 1/6 kHz. The nonlinear signal is $E_{AB}^{EO} - E_A^{EO} - E_B^{EO}$, where $E_{AB}^{EO}$ represents the electro-optic response when both excitation pulses are present, and $E_A^{EO}$ and $E_B^{EO}$ are the electro-optic responses with each pulse present individually.

In Figure~\ref{Fig2}(b), the time profiles of both excitation pulses and the generated nonlinear signal are shown for three inter-pulse time delays $\tau$ = 0, 0.5, and 1 ps. We now point out two salient features of these measurements. First, a nonlinear signal is generated only when the two excitation pulses are overlapping in time, and it moves backwards in time with the overlap region as $\tau$ increases. Second, the time profile of the nonlinear signal at the delays shown is clearly unipolar, which reflects rectification arising from a second-order coordinate response of the ZnTe \cite{Sell2008}. We remark that this signal is clearly not a radiated field, but is essentially the electro-optic Kerr effect \cite{Boyd_Book} driven by two distinct excitation fields. The full two-dimensional time evolution of the nonlinear signal is shown in Figure~\ref{Fig2}(c), which primarily exhibits the unipolar signal at early delays $\tau$ and an interesting sub-structure beyond $\tau = 1$ ps.

\section{Phenomenological model}

\begin{figure}[H]
    \centering
    \includegraphics[width=0.45\textwidth]{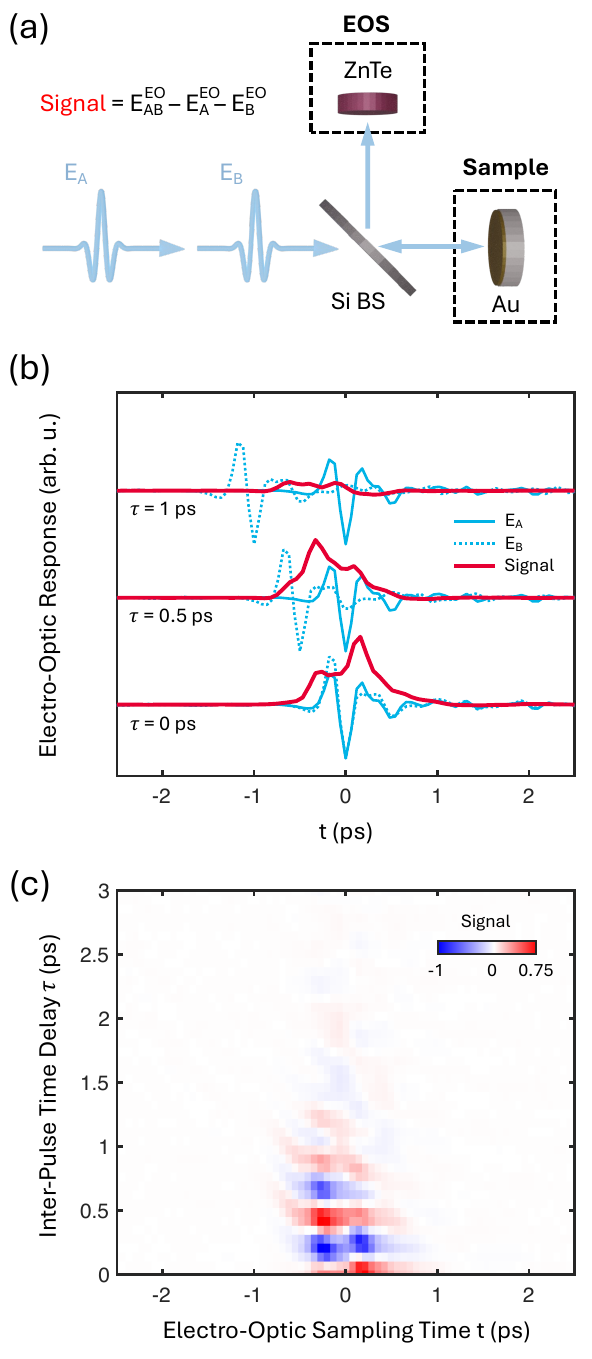}
    \caption{(a) 2DTS experiment in a reflection geometry, with a gold mirror placed at the sample position. (b) Excitation fields $E_{A/B}$ plotted alongside the non-resonant nonlinear signal for three time delays as indicated. (c) Two-dimensional time profile of the nonlinear signal as a function of inter-pulse time delay $\tau$ and laboratory time $t$.}
    \label{Fig2}
\end{figure}

\begin{figure*}
    \centering
    \includegraphics[width=1\textwidth]{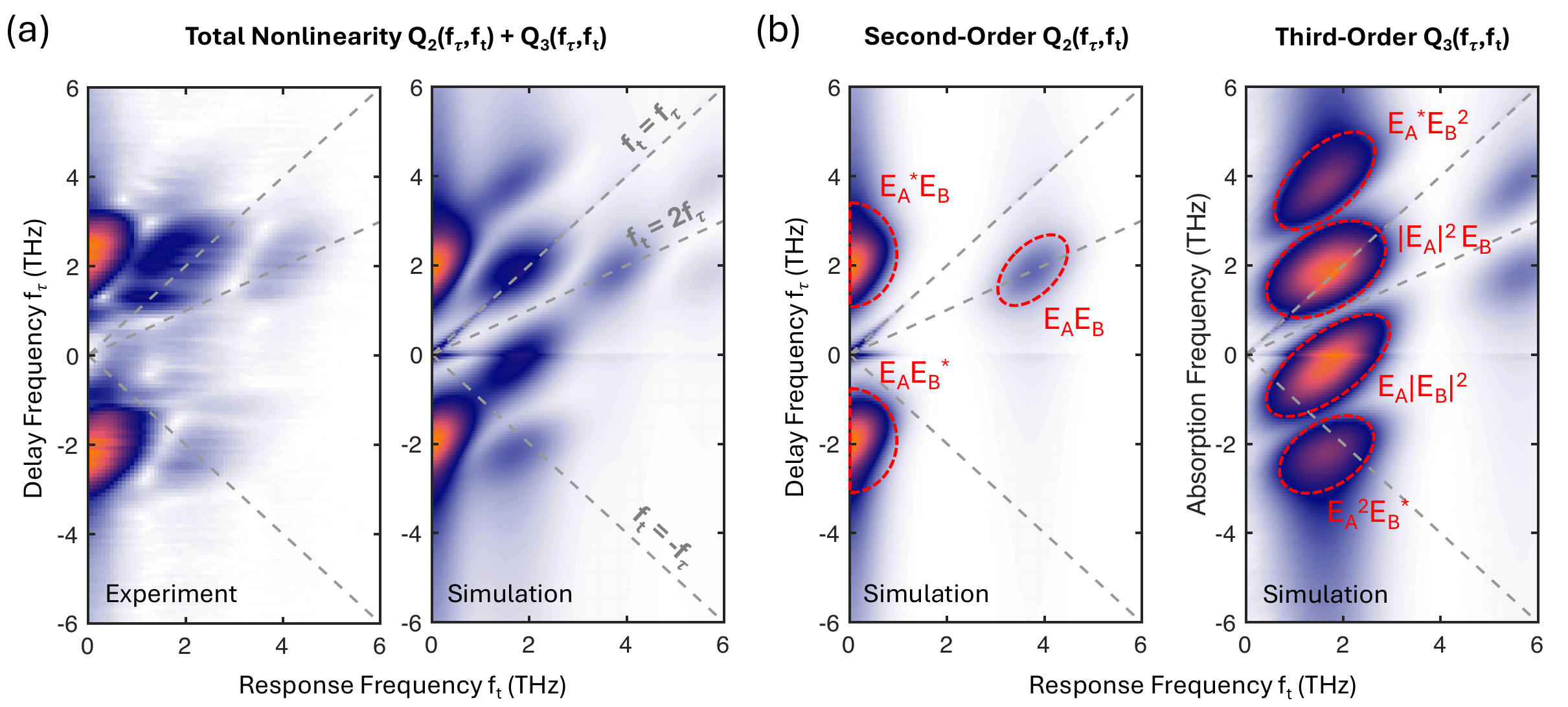}
    \caption{(a) Comparison of the experimental 2DTS spectrum of ZnTe and a corresponding spectrum simulated as described in the text. Both are acquired in an off-resonant excitation condition. (b) Spectra of solely the second-order and third-order coordinate responses as indicated, which correspond to the electro-optic Kerr effect and fourth-order electro-optic effect respectively. Each peak is labeled by its underlying wave-mixing term.}
    \label{Fig3}
\end{figure*}

To understand the complex dynamics exhibited in Figure~\ref{Fig2}(c), we Fourier transform along both time axes $\{\tau,t\}$, obtaining a two-dimensional spectrum. The resultant spectrum is shown in the left panel of Figure~\ref{Fig3}(a), which exhibits multiple peaks that correspond to distinct electro-optic nonlinearities of the ZnTe.

The origin of each feature can be understood by considering a simple model in which a nonlinear oscillator (e.g. a phonon) is driven (in this case, non-resonantly) by the electric field $E(t)$. The total energy of such a system can be written as:
\begin{align}
    H = H_K + H_U + H_{L-M}
\end{align}
where we have a kinetic energy, $H_K = \frac{1}{2}m\left(\frac{dQ}{dt}\right)^2$, and a light-matter coupling term $H_{L-M} = -QE(t)$ where $Q$ is the oscillator coordinate and $m$ its effective mass. The potential energy $H_U$ contains a harmonic term with a natural frequency $\omega_0$ in addition to anharmonic contributions. Since ZnTe lacks inversion symmetry, we include both cubic and quartic anharmonicities:
\begin{align} \label{Potential}
    H_U = \frac{1}{2}m\omega_0^2Q^2 + \frac{1}{3}maQ^3 + \frac{1}{4}mbQ^4
\end{align}
For $a > 0$, the potential stiffens with increasing excursions of $Q$ and is therefore the usual case to satisfy stability criteria.

The equation of motion for $Q(t)$ is given by:
\begin{align}\label{FullEquationOfMotion}
    \nonumber \frac{d^2Q}{dt^2} &= -\frac{1}{m}\frac{\partial H}{\partial Q}\\
    &= -\omega_0^2Q - aQ^2 - bQ^3 + E(t) - \gamma\frac{dQ}{dt},
\end{align}

where we've also included a phenomenological damping term with a damping rate $\gamma$. The above equation may be solved perturbatively by assuming an ansatz for $Q(t)$ as a power series in $\lambda$, a dummy variable that we can set to unity at the end of our analysis:
\begin{align}
    Q(t) = \lambda Q_1(t) + \lambda^2Q_2(t) + \lambda^3Q_3(t) + \dots
\end{align}
Plugging this solution into (\ref{FullEquationOfMotion}) and collecting terms in powers of $\lambda$, we obtain a hierarchy of equations:
\begin{subequations} \label{EOM}
\begin{align} \label{Q1}
    &\ddot{Q}_1 = -\omega_0^2Q_1 - \gamma\dot{Q}_1 + E(t)\\ \label{Q2}
    &\ddot{Q}_2 = -\omega_0^2Q_2 - \gamma\dot{Q}_2 - aQ_1^2\\ \label{Q3}
    &\ddot{Q}_3 = -\omega_0^2Q_3 - \gamma\dot{Q}_3 - aQ_1Q_2 - bQ_1^3\\
    \nonumber &\vdots
\end{align}
\end{subequations}
Note that, from the simple cubic and quartic nonlinearities in $H_U$, a cascade of nonlinearities follow to infinite order. The corrections $Q_n(t)$ decrease in importance with increasing order $n$, so we truncate our solution at $Q_3(t)$.

Equations (\ref{EOM}) provide an intuitive picture for the nonlinear dynamics upon driving. Equation (\ref{Q1}) for $Q_1(t)$ describes a harmonic oscillator of resonance frequency $\omega_0$ driven by $E(t)$, and therefore corresponds to solely the linear response of $Q(t)$. Equations (\ref{Q2}) and (\ref{Q3}) then each describe a harmonic oscillator of identical resonance frequency $\omega_0$ but driven indirectly by $-aQ_1^2$ and $-aQ_1Q_2 -bQ_1^3$ respectively. The nonlinear signal is then determined by a combination of $Q_2(t)$ and $Q_3(t)$, and in principle higher-order contributions that are assumed to be negligible.

We now numerically solve the above equations of motion with two identical excitation pulses of carrier frequency $\omega_c = 2$ THz and pulse duration $\sigma_t = 400$ fs that resemble the experimental pulse spectra. To simulate the non-resonant excitation condition, the oscillator frequency is set to the transverse-optical phonon frequency of ZnTe with $\omega_0 = 5.3$ THz and a large damping rate of $\gamma = 285$ THz to preclude any resonant effects. The result of simulating the total nonlinearity $Q_2 + Q_3$ is shown in the right panel of Figure~\ref{Fig3}(a), which reproduces nearly all of the salient peak positions and lineshapes of the experimental spectrum. 

We can further clarify the origins of each peak by plotting the second- and third-order coordinate responses individually, which are shown in Figure~\ref{Fig3}(b). To then identify the underlying wave-mixing combinations, we can analyze the equations of motion heuristically by noting that $Q_1$ will yield a general solution of the form:
\begin{align}
    \nonumber Q_1 &= Q_1^A\left[E_Ae^{i\omega_ct} + E_A^*e^{-i\omega_ct}\right]\\ 
    &\hspace{2cm}+ Q_1^B\left[E_Be^{i\omega_c(t+\tau)} + E_B^*e^{-i\omega_c(t+\tau)}\right]
\end{align}
where the oscillation frequency is set to $\omega_c$ due to the non-resonant excitation condition and the response amplitudes and oscillation phases are combined into the prefactors $Q_1^{A/B}$. The forcing function (dropping the terms proportional to $(Q_1^A)^2$ and $(Q_1^B)^2$ which are filtered by the detection scheme) for $Q_2$ is then directly found as $-aQ_1^2 = Q_1^AQ_1^B\left(E_AE_Be^{i\omega_c(2t + \tau)} + E_AE_B^*e^{-i\omega_c\tau} + \text{c.c.}\right)$, giving a general solution:
\begin{align}
    Q_2 = Q_2^{AB}\left[E_AE_Be^{i\omega_c(2t + \tau)} + E_AE_B^*e^{-i\omega_c\tau} + \text{c.c.}\right]
\end{align}
where relevant prefactors have once again been combined into $Q_2^{AB}$. We may now directly read the oscillation frequencies in each phase factor to determine their respective coordinates in a two-dimensional spectrum, which are $\{f_t,f_\tau\} = \{2\omega_c,\omega_c\}$ and $\{f_t,f_\tau\} = \{0,-\omega_c\}$ for the first and second terms respectively. Performing the same procedure for $Q_3$, we find:
\begin{align}
    \nonumber Q_{\text{NL}} &\propto Q_3^{A|B|^2}E_A|E_B|^2e^{i\omega_c t} + Q_3^{A^*BB}E_A^*E_B^2e^{i\omega_c(t + 2\tau)}\\
    \nonumber &\hspace{0.4cm}+ Q_3^{|A|^2B}|E_A|^2E_Be^{i\omega_c(t + \tau)} + Q_3^{AAB^*}E_A^2E_B^*e^{i\omega_c(t - \tau)}\\
    \nonumber &\hspace{0.4cm}+ Q_3^{ABB}E_AE_B^2e^{i\omega_c(3t + 2\tau)} + Q_3^{AAB}E_A^2E_Be^{i\omega_c(3t + t)}\\
    &\hspace{0.4cm}+ \text{c.c.}
\end{align}
where we can likewise read the frequency coordinates, which have been labeled in Figure~\ref{Fig3}(b). Again, nonlinearities that scale with only a single excitation pulse are dropped due to the detection scheme. We note that the peaks located at $\{f_t,f_\tau\} = \{2,4\}$ THz and $\{2,0\}$ THz are not pronounced in the experimental spectra, most likely due to the lower peak field of $E_A$ with respect to that of $E_B$ and the quadratic scaling of these two peaks with $E_A$.

\section{Discussion}

\begin{figure}[t]
    \centering
    \includegraphics[width=0.5\textwidth]{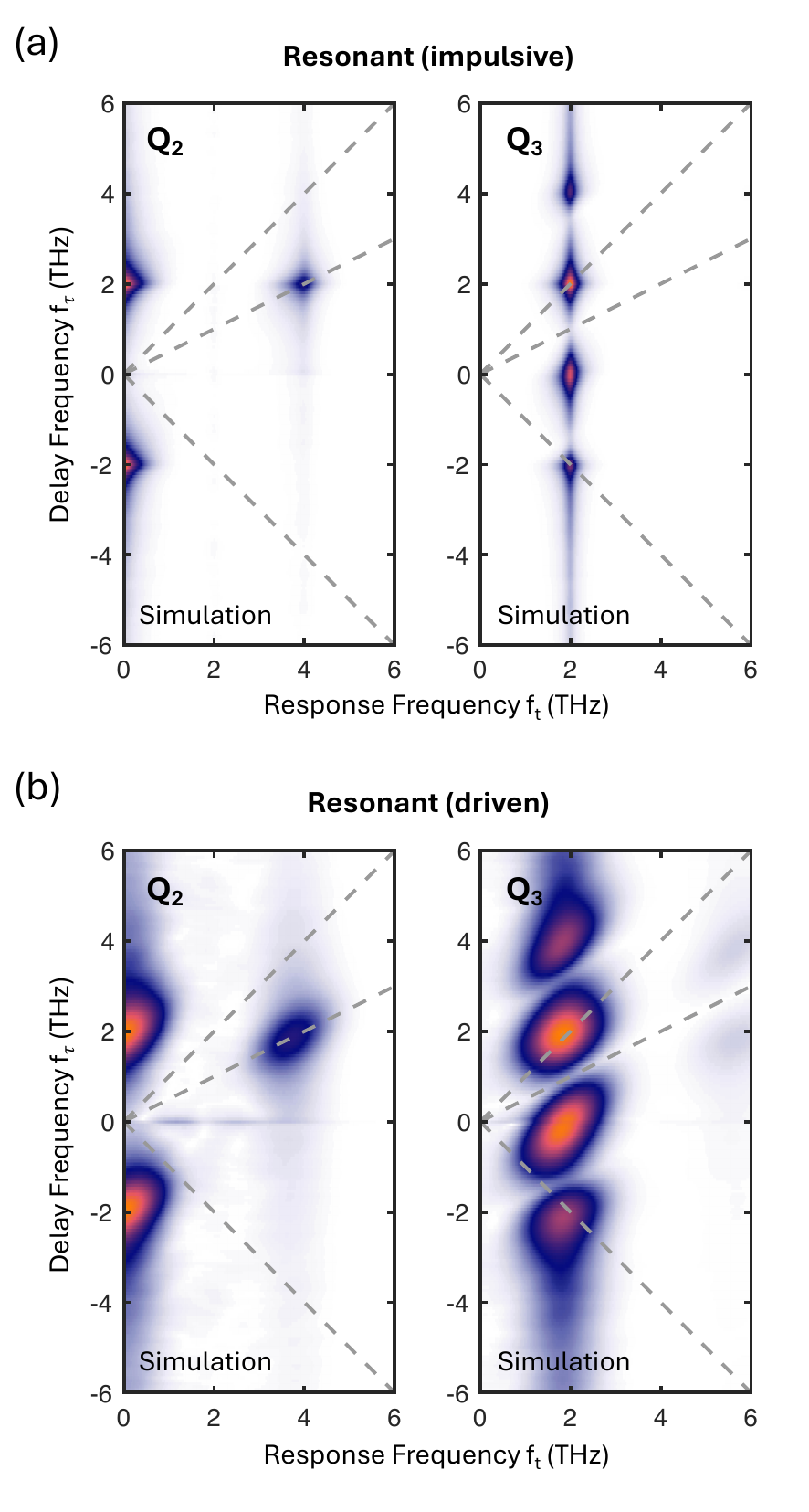}
    \caption{Simulated 2DTS spectra under resonant driving $(\omega_0 = \omega_c = 2$ THz) in the (a) impulsive limit ($\gamma = 1$ THz) and (b) driven limit ($\gamma = 10$ THz). In the impulsive limit, symmetric peak lineshapes are observed that resemble those from conventional multidimensional spectra, while in the driven limit the spectra resemble the non-resonant spectra presented in Figure~\ref{Fig3}.}
    \label{Fig4}
\end{figure}

The model presented above matches the experimental spectra well, showing the unique features of non-resonantly driven nonlinear systems. With this understanding, we now show that equivalent features can also arise from non-impulsive, resonantly driven systems (i.e. when the incident THz field bandwidths are on the order of or greater than that of the oscillator). In Figure~\ref{Fig4}(a) we show the familiar resonant spectra observed in the impulsive limit, simulated with the same parameters as in Figure~\ref{Fig3}, but with $\omega_0 = 2$ THz and $\gamma = 1$ THz. These spectra exhibit symmetric lineshapes along both axes $f_\tau$ and $f_t$. In Figure~\ref{Fig4}(b), the damping rate is increased to $\gamma = 10$ THz to exceed the excitation spectral bandwidth, resulting in a nonlinear signal generated solely during temporal overlap of the excitation pulses. The spectra in Figure~\ref{Fig4}(b) bear a remarkable resemblance to the non-resonant spectra shown in Figure~\ref{Fig3}(b), since in both cases the nonlinear response is predominately shaped by the excitation spectrum.

Indeed, a distinctive feature of the spectra in Figure~\ref{Fig3}(b) and Figure~\ref{Fig4}(c) are their asymmetric lineshapes, as all peaks exhibit a marked elongation along the $f_\tau = f_t$ direction (from the origin to the upper right corner). This may be understood by considering the fact that the temporal overlap of the pulses, during which a non-resonant (or non-impulsive) nonlinear signal is generated, moves backwards in time as the inter-pulse time delay $\tau$ increases (which can be seen for example in Figure~\ref{Fig2}). The frequency-domain representation of such a non-resonant signal along $t = -\tau$ may be most easily intuited by comparison to a photon echo signal along $t = \tau$, which results in the inverse spectral lineshape along $f_\tau = -f_t$.

As noted above, the experimental setup used here is deliberately generic. Replacing the gold mirror in Figure~\ref{Fig2}(a) with any material of interest results in a typical collinear 2DTS experiment, measured in reflection. In such an experiment, ideally one only measures the emitted nonlinear signal ($E_{AB}^{EO} - E_A^{EO} - E_B^{EO}$) generated from the mixing of $E_A$ and $E_B$ in the sample. If only the nonlinear signal is present in the EOS crystal (e.g. ZnTe), one can measure its field profile in time through the standard linear electro-optic (Pockels) effect. The results presented here, though, reveal possible artifacts (similar to the well-known `coherent artifacts' in transient absorption spectroscopy \cite{Lorenc2002}) that may arise in such a geometry in which, in addition to the nonlinear signal, both excitation pulses, $E_A$ and $E_B$ reflect off the sample and reach the EOS crystal. Therefore, even with a double-chopping scheme, a non-resonant nonlinearity of the form presented here may be observed in the signal from the EOS crystal, in addition to the material response of interest. Care must therefore be taken in collinear experiments to verify the absence of rectified nonlinearities that scale as $E_A^*E_B$ or $E_AE_B^*$, which are the leading terms in the quadratic electro-optic nonlinearity from the EOS crystal. These terms lead to responses at zero frequency and, thus, should not radiate from a probed sample. Such zero frequency signals present in the measured nonlinear response are, therefore, artifacts of the detection scheme in a collinear geometry. To avoid nonlinear excitation of the EOS crystal entirely, one can implement a non-collinear geometry, in which only the nonlinear electric field from the sample is able to reach the EOS crystal \cite{Liu_2023_echo}.

\section{Conclusion}
In conclusion, we have performed two-dimensional terahertz spectroscopy in a non-resonant excitation condition on ZnTe. The resultant spectrum exhibits unique features and lineshapes that are reproduced well by a simple simulation. We also present a perturbative analysis of the equations of motion to assign wave-mixing terms to each observed peak, and point out the lineshape elongation that may be used as indication of a spectrum taken in either a driven or non-resonant excitation condition. Finally, our results reveal potential nonlinear electro-optic artifacts that may contaminate spectra taken in collinear experimental geometries.

\end{document}